\documentclass{aa}

\usepackage{graphicx}
\usepackage{txfonts}
\usepackage{lipsum}
\usepackage{natbib}
\usepackage{subcaption}
\usepackage{lscape}
\usepackage{placeins}
\usepackage{microtype}

\newcommand{\hi}{\mbox{H\,{\sc i}} }

\begin{document}

   \title{Merger remnant in TXS 1033+026 -- I}

   \subtitle{$\rm H\alpha$ and \hi 21 cm observations suggest counter-rotating gas}

\author{J. N. H. S. Aditya\inst{1,2,3}\fnmsep\thanks{Corresponding author: adi.jnhs@shao.ac.cn, adi.jnhs@gmail.com }
        \and Hyein Yoon \inst{4,2}
        \and  L. G. Dahmer-Hahn \inst{1,5}
        \and Renzhi Su \inst{1}
        \and Rajeshwari Dutta \inst{6}
        \and Xiaolong Yang \inst{1}
        \and Elaine Sadler \inst{2,7,8}
        \and Matthew T. Whiting \inst{7}
        }

  \institute{Shanghai Astronomical Observatory, CAS, 80 Nandan Road, Shanghai 200030, P. R. China
   \and Sydney Institute for Astronomy, School of Physics A28, University of Sydney, NSW 2006, Australia
       \and ARC Centre of Excellence for All-Sky Astrophysics in 3 Dimensions (ASTRO 3D)
    \and Korea Astronomy and Space Science Institute, 776 Daedeok-daero, Yuseong-gu, Daejeon 34055, Republic of Korea
    \and Institute of Cosmology and Gravitation, University of Portsmouth, 1-8 Burnaby Road, Portsmouth, PO1 3FX, UK
    \and IUCAA, Postbag 4, Ganeshkhind, Pune 411007, India
    \and ATNF, CSIRO Space \& Astronomy, Australia
    \and ARC Centre of Excellence for Gravitational Wave Discovery (OzGrav), Australia
    }

   \date{Received September 30, 20XX}

  \abstract
   {We conducted eMERLIN and Gemini integral field unit (IFU) observations towards TXS 1033+026, a post-merger ultraluminous infrared galaxy, to study the radio continuum and $\rm H\alpha$ and \hi 21 cm lines at sub-kiloparsec scales. We find a symmetric, two-sided core-jet structure in both L- ($\approx 1.4$ GHz) and C- ($\approx 4.8$ GHz) band continuum images. The active galactic nucleus likely has a double-double radio morphology, indicating possible multiple episodes of activity. The Gemini IFU observations provided sub-arcsecond resolution. Results from these observations suggest the presence of a kiloparsec-scale circumnuclear, warped $\rm H\alpha$ disc that rotates around the core. The \hi 21 cm maps reveal a complex kinematic structure, with indications of two counter-rotating gas components orbiting the core-jet axis; one of these components counter-rotates relative to the kiloparsec-scale $\rm H\alpha$ disc. However, the true morphology of this structure may be more complex than our interpretation suggests. We estimate a \hi 21 cm velocity-integrated optical depth of $\rm 90 \pm 3 \ km \ s^{-1}$. We note that these are the first results to probe $\rm H\alpha$ and cold neutral hydrogen gas in TXS 1033+026 at sub-kiloparsec scales.
   }

\keywords{Galaxies: active -- Galaxies: ISM --
 Radio lines: galaxies
-- Methods: observational -- Techniques: high angular resolution }

\maketitle
\nolinenumbers

\section{Introduction}

Gaseous structures at sub-kiloparsec scales are the critical link between the interstellar medium and the central active galactic nucleus (AGN). These structures funnel material towards the nucleus, fuelling accretion, and are shaped by interactions with expanding AGN jets \citep[e.g.][]{gaspari2013, morganti2023}. During galaxy mergers, tidal forces and gravitational torques drive large amounts of gas to accumulate in reservoirs at these scales. This could lead to the formation of sub-kiloparsec structures such as circumnuclear discs, dense nuclear rings, and clumpy starbursts \citep[e.g.][]{bekki1995, calabro2019, hopkins2013, ueda2014}.

TXS 1033+026 is an ultraluminous infrared galaxy (\citealp[][]{kilerci2014}) and a post-merger system in which the galaxies are in the final stages of interaction \citep[][]{ellison2013}. A tidal tail, likely formed as a result of the merging activity, can also be seen connected to the system in optical images \citep[][]{dutta2018}. \citet[][]{best2012} report a redshift of $z=0.0504$, which yields a scale of 0.991 kpc arcsec$^{-1}$. \citet[][]{marecki2006} and \citet[][]{dutta2018} reported two strong optical peaks in the Sloan Digital Sky Survey r-band image (\citealp[][]{abolfathi2017}) separated by $\approx 2$ kpc. \citet[][]{dutta2018} found that the two peaks are spatially resolved in their Southern African Large Telescope (SALT) observations and detected a number of metal emission and absorption lines associated with both peaks. Based on the classification of the system as a galaxy merger, \citet[][]{dutta2018} found that the two peaks probably correspond to two interacting nuclei.

Using Giant Metrewave Radio Telescope (GMRT) observations towards the target, \citet[][]{dutta2018} detected strong radio emission at 1.4 GHz for a component identified as an AGN core, which is spatially closer to the northern optical peak. The core has integrated and peak flux densities of 165 mJy and 132 $\rm mJy \ beam^{-1}$, respectively. A fainter radio lobe is detected at a distance of 18 kpc from the core, towards the south-east. In addition to the core and the lobe at 18 kpc, \citet[][]{marecki2006} reported another radio lobe at a distance of $\approx 123$ kpc, north-west of the core, detected in the 1.4 GHz Faint Images of the Radio Sky at Twenty-Centimeters survey (FIRST; \citealp[][]{white1997}). It is not clear whether or not the two radio lobes originate from the core. \citet[][]{marecki2006} also reported a 5 GHz parsec-scale continuum image from the Multi-Element Radio Linked Interferometer Network (MERLIN). At parsec scales, the GMRT radio core shows a complex nuclear region, along with several diffuse and faint components spread along the south-east to north-west direction.

\citet[][]{dutta2018} detected strong \hi 21 cm absorption towards the core, which has a full width at zero intensity of $\rm \approx 600 \ km \ s^{-1}$. The absorption profile is fitted with three Gaussian components, with full widths at half maximum (FWHMs) of $\rm 224 \ km \ s^{-1}$, $\rm 84 \ km \ s^{-1}$, and $\rm 140 \ km \ s^{-1}$. The velocity difference between the peaks of the two strongest absorption components is $\rm \approx 150 \ km \ s^{-1}$, which is comparable to the velocity difference between the two optical peaks estimated using their redshifts \citep[][]{dutta2018}. However, \citet[][]{dutta2018} detected only a single radio emission component, the core, in the vicinity of the two optical peaks. Note that the resolution of the GMRT image is $\approx 2''$, which is comparable to the spatial separation between the optical peaks. Therefore, it is not clear whether two radio AGNs are embedded within the radio core or whether all of the \hi 21 cm absorption arises from the same or different radio AGNs.

In this work, we report results from our enhanced Multi-Element Remotely Linked Interferometer Network (eMERLIN; 1.4 GHz and 5 GHz) and Gemini Multi-Object Spectrograph North (GMOS-N) integral field unit (IFU) observations. We investigated this system to (a) identify the structure of the radio source(s) at sub-kiloparsec scales and (b) understand the gaseous structure surrounding the AGN at these scales by tracing \hi 21 cm absorption and $\rm H\alpha$ emission lines. These are the first sub-kiloparsec-scale observations of \hi 21 cm absorption and $\rm H\alpha$ emission towards this object. In the following, we describe the radio and optical observations, data reduction, and interpretation of the observational results.
\section{Observations and data reduction} \label{sec:observations}

\subsection{eMERLIN observations}

We conducted eMERLIN observations towards TXS 1033+026 in the L ($\rm \approx 1.4 \ GHz$) and C ($\rm \approx 4.8 \ GHz$) bands (P.I. J.N.H.S. Aditya). The L-band observations were conducted on 18--19 December 2023, whereas the C-band observations were conducted on 9 January 2023. The C-band observations have a bandwidth of 128 MHz, centred at 4.77 GHz, and divided into 128 channels. The on-source time was 2 h. The L-band observations have a bandwidth of 16 MHz, subdivided into 512 channels (each of 31.25 kHz), covering frequencies from 1346.5 MHz to 1362.5 MHz. The channels provided a velocity resolution of $\rm 6.9 \ km \ s^{-1}$ and a coverage of $\rm 3400 \ km \ s^{-1}$ around the redshifted \hi 21 cm line. The on-source time was 12 h. In both observations, we used 3C286 for fringe-finding, bandpass and amplitude calibration, and 1028+0255 for phase-referencing.

The data were reduced using Astronomical Image Processing System (\textsc{AIPS}), following standard techniques as described below. First, the instrumental delay was corrected using a two-minute scan on the fringe-finder, 3C286. The bandpass was calibrated using scans on 3C286. Then, frequency- and time-dependent phase calibration was performed using the scans on the phase-referencing calibrator and the fringe-finder. The solutions were then transferred to the target. Then, multiple rounds of imaging and self-calibration were performed on the target until the image noise no longer improved. The self-calibration steps involved a few rounds of phase-only self-calibration, followed by one round of amplitude and phase self-calibration. Finally, a continuum image was produced in the C band, while a continuum image and a \hi 21 cm spectral cube were produced in the L band. The \textsc{Cvel} task, which includes heliocentric velocity correction, was used to estimate velocities with respect to the systemic redshift (discussed in Sect.~\ref{sec:post_processing}). The integrated flux densities of the source in the C and L bands are 61 mJy and 156 mJy, respectively. The uncertainty in the flux density is $\approx 10\%$ due to calibration errors. In addition, we extracted \hi 21 cm moment maps and spectra from the spectral cube.

The synthesised beams in the C- and L-band images have sizes of $0.25'' \times 0.09''$ and $0.36'' \times 0.19''$, respectively.
The continuum noise in the C- and L-band images is 63.5 $\rm \mu Jy \ beam^{-1}$ and 289.7 $\rm \mu Jy \ beam^{-1}$, respectively. The root mean square (RMS) noise in the spectral cube is $\rm 1.1 \ mJy \ beam^{-1}$ per 6.9 $\rm km \ s^{-1}$ channel.

\subsection{GMOS-North observations}

\begin{figure*}[!ht]

\includegraphics[width=0.49\textwidth]{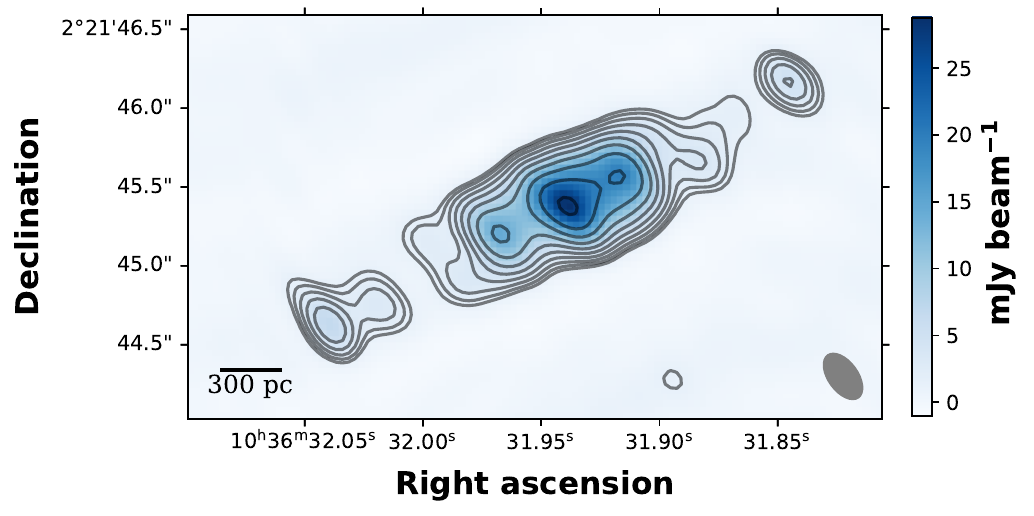}
\includegraphics[width=0.49\textwidth]{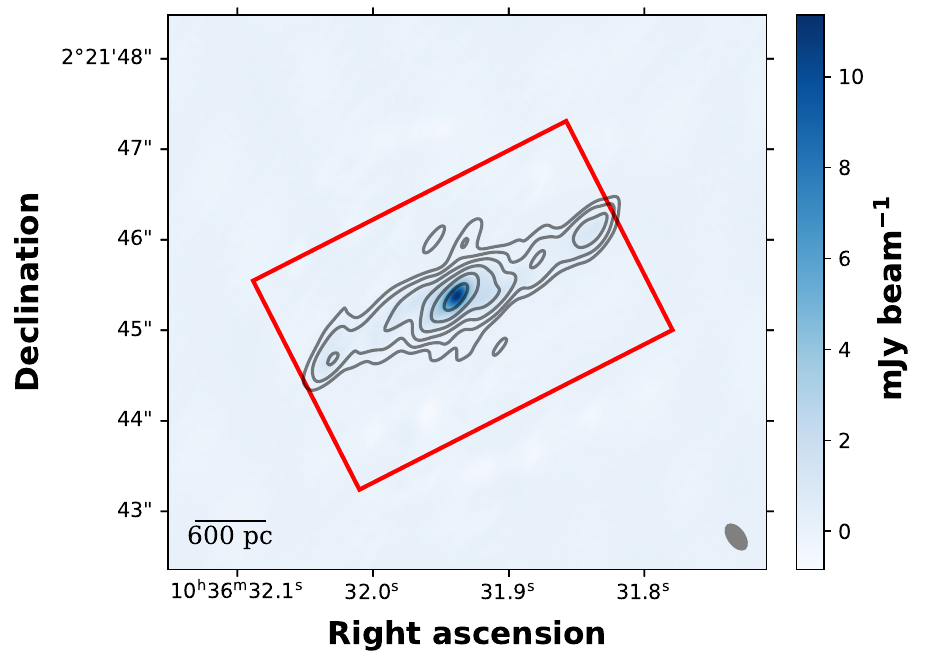}

\caption{\textit{Left}: eMERLIN L-band continuum image. The contour levels are 1.1 mJy beam$^{-1}$ $\times$ (1, $\sqrt{2}$, 2, 2$\sqrt{2}$, 4, 4$\sqrt{2}$, 8, 8$\sqrt{2}$, 16, 16$\sqrt{2}$). The peak flux density is 28.8 mJy beam$^{-1}$.
\textit{Right}: eMERLIN C-band continuum image. The contour levels are 0.35 mJy beam$^{-1}$ $\times$ (1, 2, 4, 8, 16, 32, 64). The peak flux density is 11.4 mJy beam$^{-1}$. The red rectangle represents the field of view for the Gemini IFU observations. The synthesised beam is shown in the bottom-right corner of each panel.}
\label{fig:21cm_cont_maps}
\end{figure*}

The IFU data were acquired on 25 May 2025 using the GMOS-N instrument, under project number GN-2025A-FT-105 (P.I. Hyein Yoon). We took four science frames, each with an integration time of 800 s (total time of 53 min), using the R831/G5302 grating with central wavelengths of 0.595 and 0.605 $\mu$m. Calibration images were taken on the same or subsequent days. The mean airmass for the observations is 1.15. The data covered wavelengths in the range 4761--7231 \AA. The field of view of the lenslet array has dimensions of $3.97'' \times 2.59''$. For positioning the lenslet array, we used the location of the radio peak in the C-band image as a reference and centred the array at this location. The array was placed at a position angle of $123^\circ$ (east of north), which covered nearly the entire emission in the C band (see the right panel of Fig.~\ref{fig:21cm_cont_maps}).

Data reduction was performed with standard \textsc{Iraf Gemini} packages\footnote{Following the tutorial at \url{https://gmos-ifu-1-data-reduction-tutorial-gemini-iraf.readthedocs.io/en/latest/}}. The main steps were bias subtraction, fibre tracing, wavelength calibration, flat-field correction, cosmic-ray removal, quantum efficiency correction, science data extraction, sky subtraction, and heliocentric velocity correction. For sky subtraction, a separate sky field was observed simultaneously with the science exposures. This sky frame was then used for sky subtraction before constructing the IFU datacube. We also observed a standard star, which was reduced in the same way as the science frames and later used to perform sensitivity correction and flux calibration. Lastly, different exposures with different central wavelengths were combined to remove the gaps between the CCDs. The final datacube has a seeing-limited spatial resolution of $0.6''$ (the size of the point spread function, PSF, estimated from the standard star), sampled at a spaxel size of
$0.0809'' \times 0.0809''$, and a spectral resolution (FWHM of the instrumental profile) of $\Delta\lambda \approx 1.1$\,\AA, corresponding to $R = \lambda/\Delta\lambda = 4340$ at the blue end and $6487$ at the red end of the datacube ($\Delta\lambda$ constant throughout). The datacube's spectral dispersion is $0.376$\,\AA\,pixel$^{-1}$, so the instrumental profile's FWHM spans $\approx2.9$ pixels.

\subsection{Post-processing} \label{sec:post_processing}

\begin{figure*}[!ht]
\includegraphics[width=0.5\textwidth]{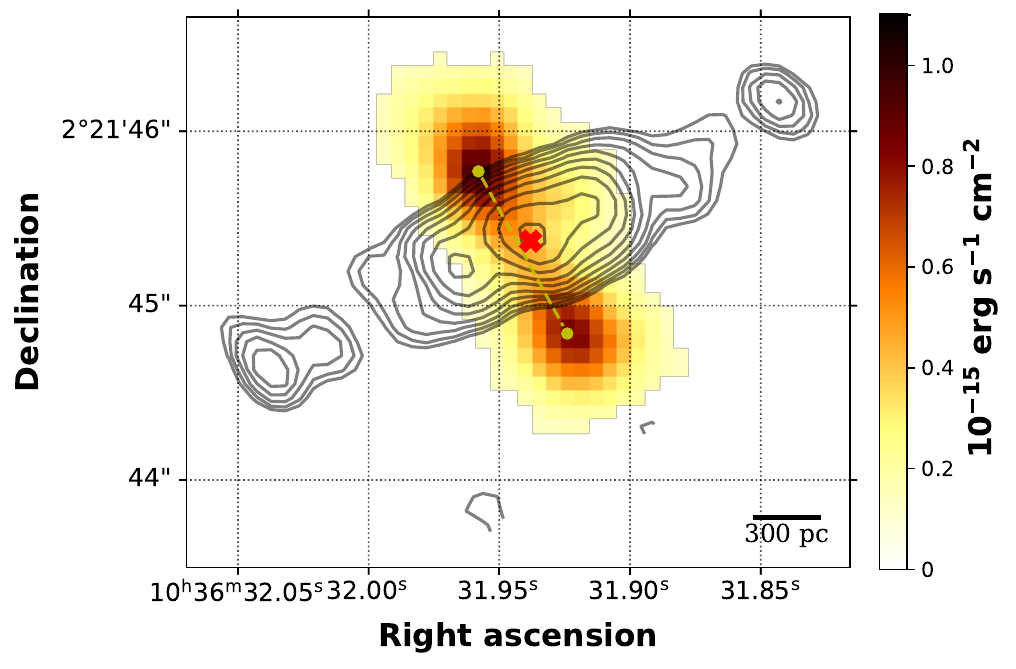}
\includegraphics[width=0.5\textwidth]{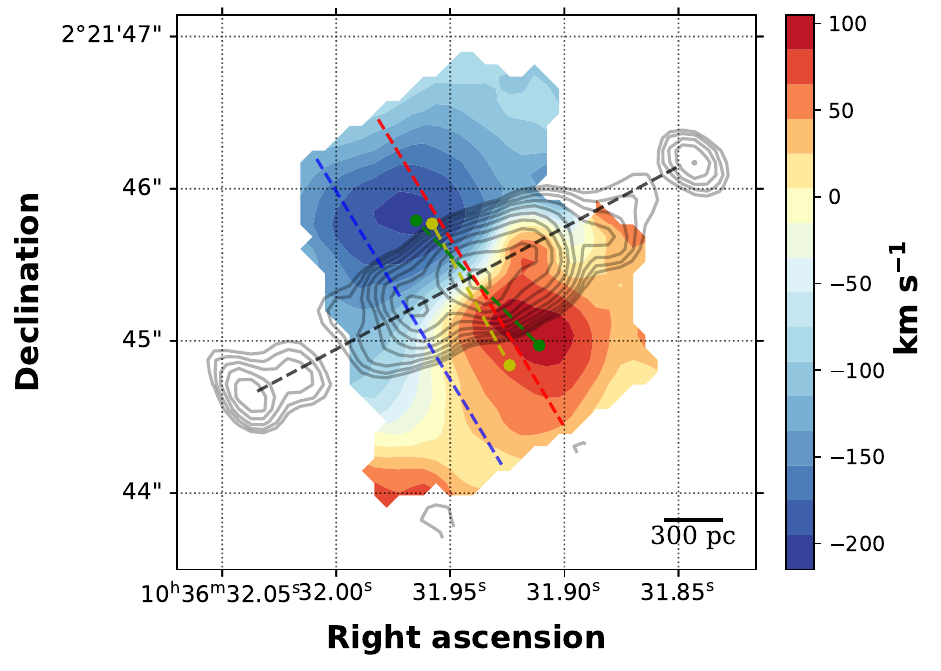}
\caption{\textit{Left}: $\rm H\alpha$ integrated flux map. The red cross represents the position of the peak emission in the C-band image. \textit{Right}: $\rm H\alpha$ line-of-sight velocity map. The velocities are relative to $z_{\rm core}$. The dashed green line connects the peak absolute velocities on both sides of the radio continuum. The dashed red and blue lines represent PV cuts A and B, respectively (see Sects.~\ref{sec:ha_disc} and \ref{sec:pv_plots}), and the dashed black line marks the core-jet axis. In both panels, the black contours represent the L-band radio continuum emission, at the same levels as those in Fig.~\ref{fig:21cm_cont_maps}, and the dashed yellow lines connect the peak optical emission on both sides of the radio continuum. }
\label{fig:HA_maps}
\end{figure*}

We extracted the optical spectrum from the reduced Gemini IFU datacube at the location of the emission peak in the C-band radio image (see Fig.~\ref{fig:HA_maps}). We fitted Gaussian profiles to the strong emission lines detected in the spectrum, $\rm H\beta$ (4861\,\AA), [O\,{\sc i}] (6300\,\AA\ and 6364\,\AA), [S\,{\sc ii}] (6716\,\AA\ and 6731\,\AA), [N\,{\sc ii}] (6548\,\AA\ and 6583\,\AA), and $\rm H\alpha$ (6563\,\AA) (see Fig.~\ref{fig:core_optical_spec}), and estimated a redshift of $z_{\rm core} = 0.05012 \pm 0.00011$. Here, three Gaussian profiles were simultaneously fitted to the two [N\,{\sc ii}] lines and the $\rm H\alpha$ component, and two profiles were simultaneously fitted to the [S\,{\sc ii}] (6716\,\AA\ and 6731\,\AA) lines. The Na\,{\sc i} doublet was not used for redshift estimation, as this doublet is blended with weak He\,{\sc i} emission, and a simple Gaussian fit could lead to errors in redshift estimation. The wavelength dispersion is $\approx 0.38$\,\AA\,pixel$^{-1}$ around the $\rm H\alpha$ line, corresponding to a velocity sampling of $\approx 17.2$\,km\,s$^{-1}$ per pixel.

We produced integrated flux, line-of-sight velocity, and velocity dispersion maps for $\rm H\alpha$ emission (see Figs.~\ref{fig:HA_maps} and \ref{fig:ha_mom2}) using the procedure described below. A tied triple-Gaussian model was fitted to the $\rm H\alpha$ and [N\,{\sc ii}] lines, with the [N\,{\sc ii}]\,$\lambda$6583/[N\,{\sc ii}]\,$\lambda$6548 flux ratio fixed at 2.96, and the line-of-sight velocity and velocity dispersion of all three lines tied, together with a linear continuum \citep[e.g.][]{freitas2018}. Least-squares fitting was performed using the \texttt{curve\_fit} routine from Python's \texttt{SciPy} package. The velocity map gives the fitted $\rm H\alpha$ centroid relative to the systemic velocity, and the dispersion map gives the fitted $\sigma$ with the instrumental resolution removed in quadrature. A Gaussian was not fitted to spaxels where the $\rm H\alpha$ peak flux density fell below 2\% of the maximum value found across all spaxels in the cube \citep[e.g.][]{durre2019}; this threshold excludes noisy data at large radii while retaining as much usable data as possible in the central regions. We confirmed that different flux thresholds gave the same results within the region shown in Fig.~\ref{fig:HA_maps}. Examples of the tied triple-Gaussian fits to the H$\alpha$ and [N\,{\sc ii}] lines at three locations across the region are shown in Fig.~\ref{fig:core_optical_spec}.

\begin{figure*}[!ht]
\includegraphics[width=0.49\textwidth]{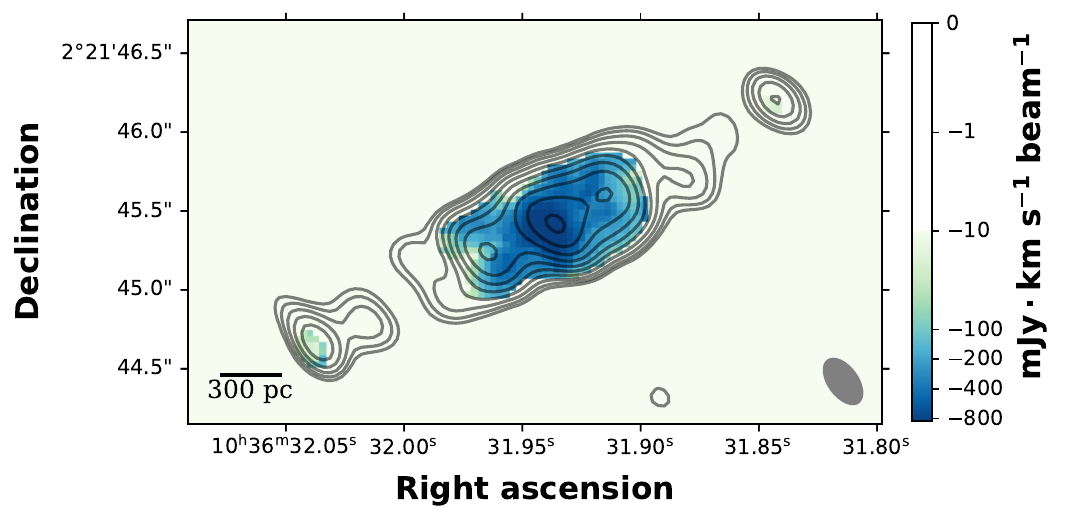}
\includegraphics[width=0.49\textwidth]{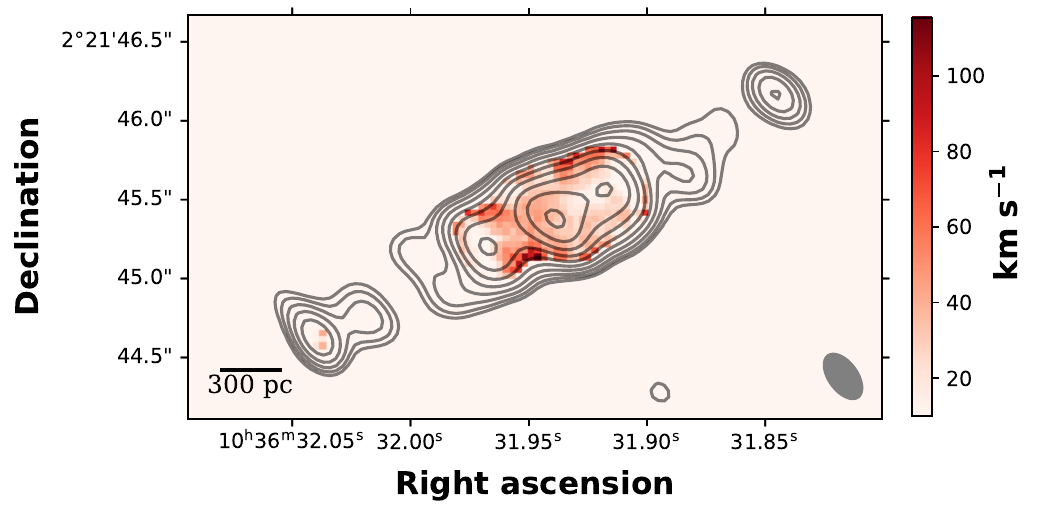}
\includegraphics[width=0.49\textwidth]{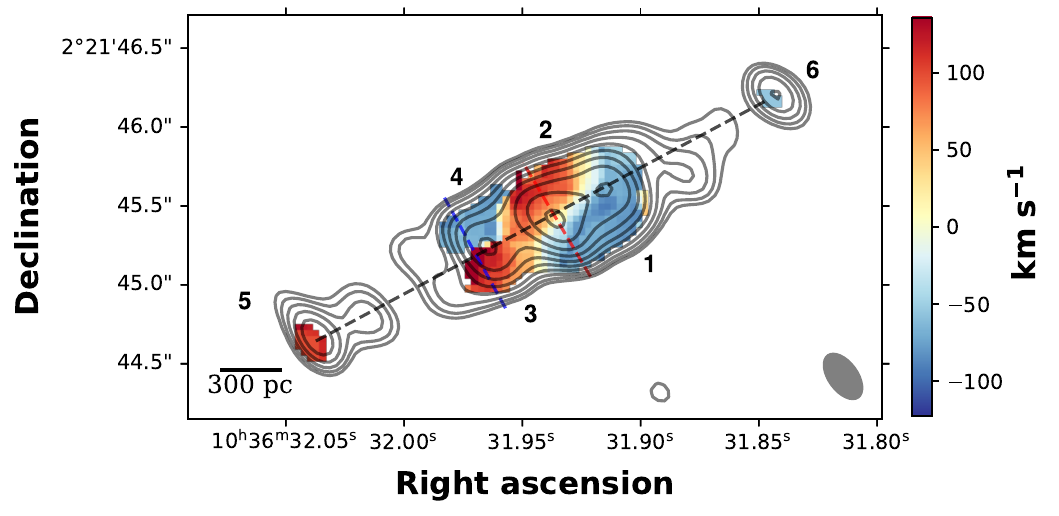}
\includegraphics[width=0.49\textwidth]{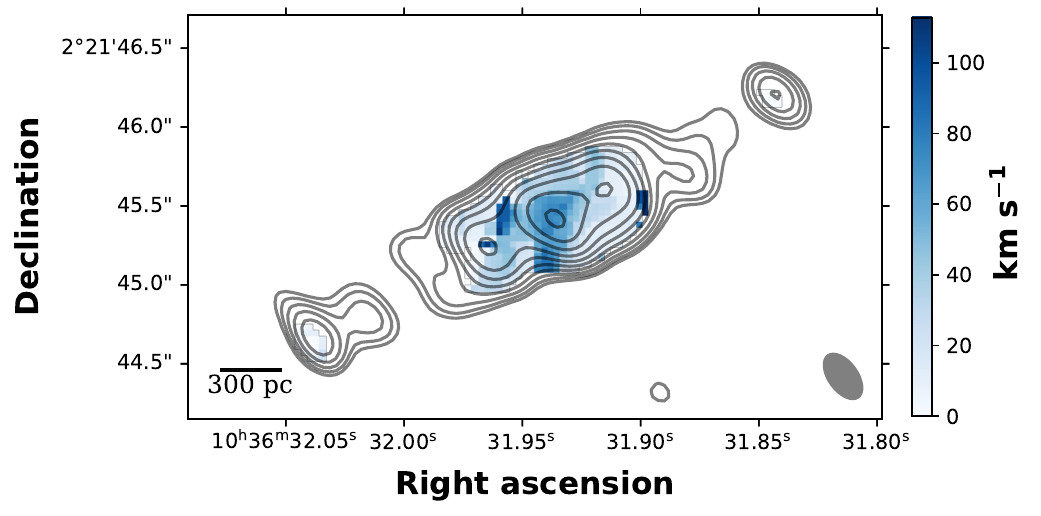}

\caption{\textit{Top left}: eMERLIN \hi 21 cm absorption total absorbed intensity (mom0). \textit{Top right}: VOD. \textit{Bottom left}: Velocity (mom1). The various regions are marked with numbers from 1 to 6, PV cuts A and B are shown as dashed red and blue lines, respectively, and the dashed black line marks the core-jet axis. The description for the placement of these cuts is given in Sect.~\ref{sec:pv_plots}.
  \textit{Bottom right}: Velocity dispersion (mom2). In all plots, the black contours represent the eMERLIN L-band continuum emission, at the same levels as those in Fig.~\ref{fig:21cm_cont_maps}. The velocities are with respect to $z_{\rm core}$. The synthesised beam is shown in the bottom-right corner of each panel.}
\label{fig:21cm_mom_maps}
\end{figure*}

Using the eMERLIN radio datacube, we extracted the \hi 21 cm absorption spectra (see Figs.~\ref{fig:tot_h1_spec} and \ref{fig:h1_spec}) and produced moment 0, 1, and 2 maps, as well as a velocity-integrated optical depth (VOD) map (see Fig.~\ref{fig:21cm_mom_maps}). The spectra and maps were produced using \textsc{AIPS}. We used a mask to include all absorption beyond $3.5\sigma$ in the line channels from the region co-spatial with the radio source \citep[see e.g.][]{murthy2024}. To reduce Gibbs ringing in the spectrum and to improve the signal-to-noise ratio (S/N) in the moment maps, we smoothed the spectra to a velocity resolution of $\rm 20.7 \ km \ s^{-1}$, and the images to a spatial resolution of $\approx 110$ pc, using Gaussian kernels. Note that the smoothed velocity channel is significantly smaller than the FWHM of the narrowest absorption component ($\rm 84 \ km \ s^{-1}$) detected by \citet[][]{dutta2018}, and the spatial resolution is significantly smaller than the L-band beam size, $\rm 356\ pc \times 186\ pc$. Therefore, the smoothing retained narrow absorption components while improving the S/N. We used $z_{\rm core}$ to estimate the systemic velocity.

Finally, we produced position--velocity (PV) maps for both $\rm H\alpha$ and \hi 21 cm lines along two cuts in the respective velocity maps. The cuts and the PV plots are described in the following sections.

\section{Structure interpretation}

\subsection{Radio continuum } \label{sec:jet_act}

In the eMERLIN continuum images (Fig.~\ref{fig:21cm_cont_maps}), we consistently see a prominent core (the region of peak emission) and a double-jet structure in both C- and L-band images. The extent of the radio emission is nearly symmetric on both sides of the core in both images, suggesting that the expansion of the radio jets is likely occurring in the plane of the sky, nearly perpendicular to our line of sight. The total size of the emission ($\rm \approx 3.7 \ kpc$) is similar in the two bands.

The jets extend in the south-east and north-west directions and are directionally aligned with the 18 kpc and 123 kpc radio lobes detected in the FIRST image \citep[][]{white1997, marecki2006}, indicating that the large-scale radio lobes are likely associated with this source. However, due to the larger distance between the core and the lobe at 123 kpc, the association between this lobe and the source is more uncertain. In addition to the core, we can see two pairs of emission peaks in the L-band image, aligned linearly and symmetrically on both sides of the radio core, indicating a double-double radio morphology \citep[][]{saikia2009}. The likely association of large-scale radio lobes with the core and the presence of multiple radio peaks suggest multiple episodes of jet activity \citep[e.g.][]{rao2023}.

\subsection{$\rm H\alpha$ disc} \label{sec:ha_disc}
In the left panel of Fig.~\ref{fig:HA_maps}, we can see two strong lobes of $\rm H\alpha$ emission located nearly symmetrically on opposite sides of the radio core. The two lobes have similar sizes, and their emission peaks are at a similar separation from the core, indicating that the lobes are likely components of a single gas structure that is gravitationally bound to the core. The emission peaks of the two lobes are separated by $\approx 1$ kpc, smaller than the $\approx 2$ kpc separation between the optical emission peaks reported by \citet[][]{dutta2018}. This suggests that the $\rm H\alpha$ gas likely has a smaller size than the region producing the optical emission.

In the right-hand panel, the map shows a well-ordered velocity field for the two lobes, with red- and blueshifted velocities relative to the systemic redshift ($z_{\rm core}$). Several iso-velocity contours are visible within each lobe, and the transition between the red- and blueshifted regions is smooth: an S-shaped, zero-velocity profile runs along the radio core and jets, connecting the two sides. The velocity dispersion map (see Fig.~\ref{fig:ha_mom2}) shows similar values across both lobes for most of the region, except in the north-western region, where it intersects the continuum; the higher values there could result from beam smearing \citep[e.g.][]{zhou2017} or from jet-gas interaction. This ordered velocity field, together with the similar dispersion across most of the two lobes, points to a rotating gas disc \citep[e.g.][]{knapen2004, ueda2014}.

\citet{dutta2018} measured redshifts of $0.04990\pm0.00005$ and $0.05049\pm0.00005$ for the two optical peaks, corresponding to velocities of $-63\pm14$~km\,s$^{-1}$ and $+106\pm14$~km\,s$^{-1}$ relative to $z_{\rm core}$. These velocities are comparable to those on the blue- and redshifted sides of our H$\alpha$ velocity field, which spans $\approx -200$ to $+100$~km\,s$^{-1}$. The velocity difference between the two optical peaks ($\approx 170$~km\,s$^{-1}$) is smaller than that between the emission peaks of the two H$\alpha$ lobes ($\approx 270$~km\,s$^{-1}$). At the lower spatial resolution of the SALT observations, the emission from the two lobes is expected to be blended, which reduces the measured velocity separation. The two optical peaks may therefore arise largely from the blue- and redshifted sides of the H$\alpha$ disc.

In Fig.~\ref{fig:ha_pv}, we show the PV plots for the two cuts (A and B) shown in the right panel of Fig.~\ref{fig:HA_maps}. Cut A passes through the radio core and is nearly perpendicular to the core-jet axis, which is a fit to the core and four emission peaks in the L-band image (see Sect.~\ref{sec:jet_act}). The selection of cut B is based on the spatial structure of the \hi 21 cm absorption and is described in Sect.~\ref{sec:pv_plots}. In both PV plots, we can see a connected structure in which the emission peaks lie on opposite sides of the midpoint of the cut and have positive and negative velocities, respectively. This profile supports the presence of a gas disc rotating around the nucleus or around the radio-jet axis \citep[e.g.][]{ueda2014}.

Further, the line connecting the peaks of the absolute velocities of the two lobes appears to be skewed relative to the line connecting the emission intensity peaks (see the right panel of Fig.~\ref{fig:HA_maps}). The angle between the two connecting lines is $\approx 19^\circ$. The inverted S-shaped zero-velocity profile and the probable skew between total intensity and velocity peaks likely indicate the presence of a warp in the gas disc \citep[e.g.][]{ueda2014, kolcu2023}.

\subsubsection{Model fitting} \label{sec:model_fit}

To quantify the geometric properties of the $\rm H\alpha$ structure, we fitted a tilted-ring model to the velocity field. We chose a tilted-ring model because the data indicate the presence of a warp in the gas disc. This routine uses a least-squares fitting algorithm to estimate the geometric parameters, similar to the {\sc AIPS} task {\sc GAL}. We fitted four concentric tilted rings with a width of $0.43''$ each, covering radii of $0.0''$--$1.7''$ from the centre (see Fig.~\ref{fig:ring_fits}). All rings share a single, common kinematic centre, which is a free parameter of the fit. The model is PSF-convolved at each iteration using a flux-weighted Gaussian convolution, so the fitted parameters account for the PSF.
The position of the radio peak in the C-band image was provided as an initial guess for the disc centre: 10:36:31.938, +02:21:45.37 (RA, Dec; J2000). The fits converged and yielded (a) the position of the kinematic centre of the rings (RA and Dec), (b) the position angle of the receding (redshifted) major axis of each ring, measured from north to east in the map, and (c) the inclination of each ring. We have plotted the observed, model, and residual velocities for $\rm H\alpha$ in Fig.~\ref{fig:ring_fits}, and provided the fit parameters for the four rings in Table~\ref{tab:ringfit}. The fitted centre and the median position angle and inclination of the four rings are given in Table~\ref{tab:gal_fit}. The fitted centre is consistent with the C-band radio peak position, within the $0.6''$ PSF. The good agreement between the fitted tilted-ring model and the observed velocity field indicates the presence of a rotating gas disc \citep[e.g.][]{ueda2014}, and the coincidence between the fitted centre and the radio peak suggests that this gas is rotating around the radio core.

\begin{table}
\caption{Parameters of the tilted-ring fit.}
\label{tab:gal_fit}
\centering
\begin{tabular}{cccc}
\hline
RA & Dec & PA & Inc. \\
\hline
10:36:31.948 & +02:21:45.47 & 231$^\circ$ & 38$^\circ$ \\
$\pm0.6''$ & $\pm0.6''$ & $\pm12^\circ$ & $\pm11^\circ$ \\
\hline
\end{tabular}
\tablefoot{The columns are: the RA and Dec of the fitted kinematic centre of the rings, and the median position angle and inclination of the four rings. The second row shows fitting uncertainties.}
\end{table}

\subsection{\hi 21 cm absorption structure}

In Fig.~\ref{fig:21cm_mom_maps} we show the \hi 21 cm absorption moment maps (mom0, mom1, and mom2) and the VOD map.

\subsubsection{mom0 and VOD maps}
The mom0 map shows strong absorption against the core and relatively weak absorption against the two radio peaks on both sides of the core. The VOD map shows low values along the jet axis, i.e. along the core and the two radio peaks, while the VOD appears to be getting stronger towards the peripheries of the gas structure. The total absorption strength against the core, in the mom0 map, appears to be stronger than that against the adjacent radio peaks. However, that is not the case in the VOD map. This is due to the relatively strong continuum emission from the core, which causes strong absorption, rather than to a high VOD of the gas in the foreground of the core.

A low VOD along the core-jet axis could indicate either the clearance of gas along the radio peaks by the expanding jets or a projection effect of a thick gas disc rotating around the jet axis, in which the edges have higher absorption strength due to a larger column of gas intersecting the line of sight. However, it cannot be ruled out that in some of the peripheral pixels the optical depth could be higher due to spurious, noisy absorption-like features.

\subsubsection{mom1 map}
The mom1 map shows a complex velocity structure. The various regions are marked in the bottom-left panel of Fig.~\ref{fig:21cm_mom_maps}.
Each marked region is a continuous region in which the velocities are either all blueshifted or all redshifted, and which is not intercepted by a region with an absolute velocity of $\rm < 20 \ km \ s^{-1}$.

Interestingly, regions 1 and 2 have blue- and redshifted velocities with respect to $z_{\rm core}$, respectively, which are contrary to the velocities of the respective regions in the $\rm H\alpha$ velocity map (see Fig.~\ref{fig:HA_maps}). Between regions 1 and 2, we can again see an inverted S-shaped zero-velocity profile passing through the core.
On the other hand, regions 3 and 4 have red- and blueshifted velocities, respectively, which are approximately consistent with the velocities of the respective regions in the $\rm H\alpha$ velocity map. Finally, the smaller regions, 5 and 6, have red- and blueshifted velocities, respectively.

\subsubsection{PV plots} \label{sec:pv_plots}

In Fig.~\ref{fig:h1_pv_plot} we show the PV plots for cut A, which passes through regions 1 and 2, and cut B, which passes through regions 3 and 4. Cut A was chosen such that it passes through the radio core and is nearly perpendicular to the axis of the radio jets. Cut B was chosen such that it covers the maximum number of pixels in regions 3 and 4 and is nearly perpendicular to the axis of the radio jets.

In the PV plot for cut A, we can see a continuous elongated structure aligned in opposite quadrants. The profile suggests the presence of a rotating gas disc \citep[e.g.][]{ueda2014}. The orientation of the structure is opposite to that in the $\rm H\alpha$ PV plot for cut A, indicating that the \hi 21 cm gas is counter-rotating relative to the $\rm H\alpha$ disc. Note that this \hi 21 cm disc is a component of the complete absorption structure seen in Fig.~\ref{fig:21cm_mom_maps}.

In the PV plot for cut B, covering regions 3 and 4, we see two dominant disconnected structures lying in opposite quadrants. Their alignment is opposite to that in the \hi 21 cm PV plot for cut A, but similar to those in both $\rm H\alpha$ PV plots.
Since regions 3 and 4 appear to be integral components of the larger structure in the mom1 map, it is less likely that they represent simple inflowing and outflowing gas clouds. The disconnected absorption profile in the PV plot likely indicates a gas ring rotating in an opposite orientation relative to the gas in regions 1 and 2. However, the discontinuity could also be due to insufficient sensitivity of the eMERLIN data to detect any weak absorption.

\subsubsection{mom2 map}

In Fig.~\ref{fig:21cm_mom_maps}, the velocity dispersion is $\rm \lesssim 75 \ km \ s^{-1}$ in most of the region showing absorption, except at the intersections between regions 1 and 3, and regions 2 and 4, and at the north-west periphery in region 1, where the values are higher. Also, the high dispersion values appear to be confined to small regions. The localised high dispersion at the intersection regions is likely due to line broadening from beam smearing, rather than to intrinsic gas velocity dispersion. The reason for the higher dispersion at the north-west periphery is not clear; it could be due to jet-cloud or cloud-cloud interactions.

\subsection{Interpretation}

Our interpretations of the overall gas structure are as follows: (a) Regions 1 and 2 represent a gas disc rotating in an opposite sense relative to the $\rm H\alpha$ disc. There is an indication of a warp in the disc. Regions 3 and 4 represent a gas ring that is rotating in an opposite sense relative to the gas in regions 1 and 2, and nearly aligned with the orientation of rotation of the $\rm H\alpha$ disc. (b) The gas could have a complex warped and/or helical structure, creating the velocity pattern visible in the mom1 map.

The overall \hi 21 cm structure is likely part of a larger gas system. The current observations detect only the portion of the system that is in front of the background radio emission. The interpretations about the gas structure could be confirmed through further high-spatial-resolution and more sensitive observations of cold atomic and molecular gas lines.

\begin{figure}[ht]

\includegraphics[width=0.49\textwidth]{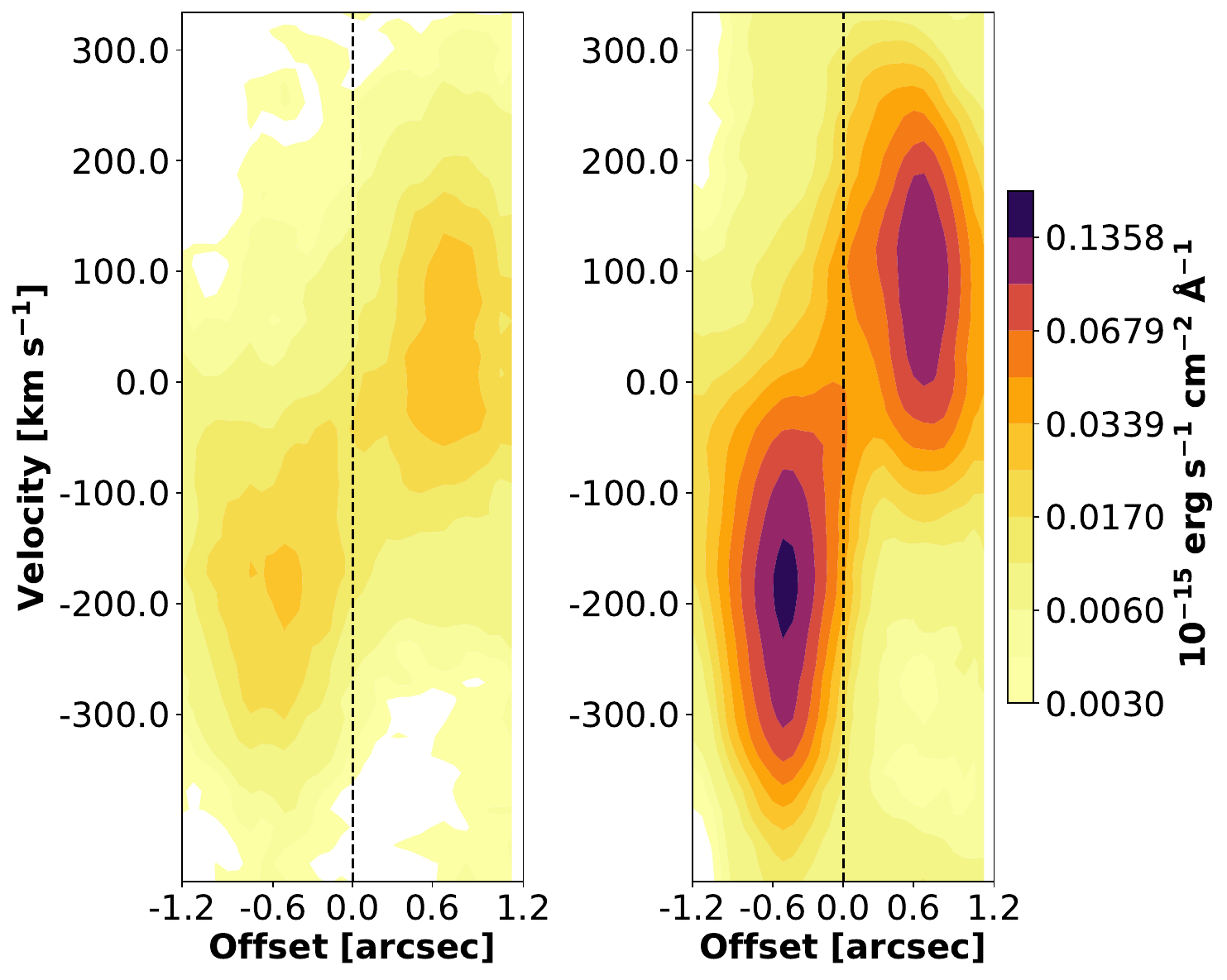}

\caption{$\rm H\alpha$ PV plots. \textit{Left}: Cut B. \textit{Right}: Cut A. The cuts are shown as dashed blue (B) and red (A) lines in the right panel of Fig.~\ref{fig:HA_maps}. The vertical dashed line represents the midpoint of the respective cuts.}
\label{fig:ha_pv}
\end{figure}

\begin{figure}[ht]

\includegraphics[width=0.49\textwidth]{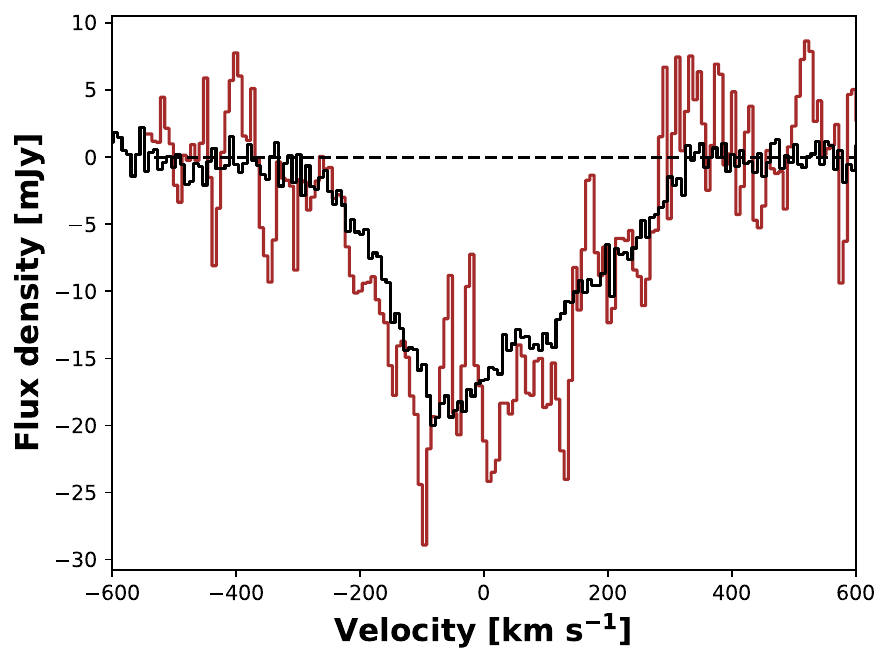}

\caption{\hi 21 cm absorption spectra from the eMERLIN and GMRT observations. The red line represents the eMERLIN \hi 21 cm spectrum for the full region where absorption is detected in Fig.~\ref{fig:21cm_mom_maps}. The black line is the GMRT \hi 21 cm absorption spectrum reported in \citet[][]{dutta2018}. The velocities are with respect to $z_{\rm core}$.}
\label{fig:tot_h1_spec}
\end{figure}

\begin{figure}[ht]

\includegraphics[width=0.49\textwidth]{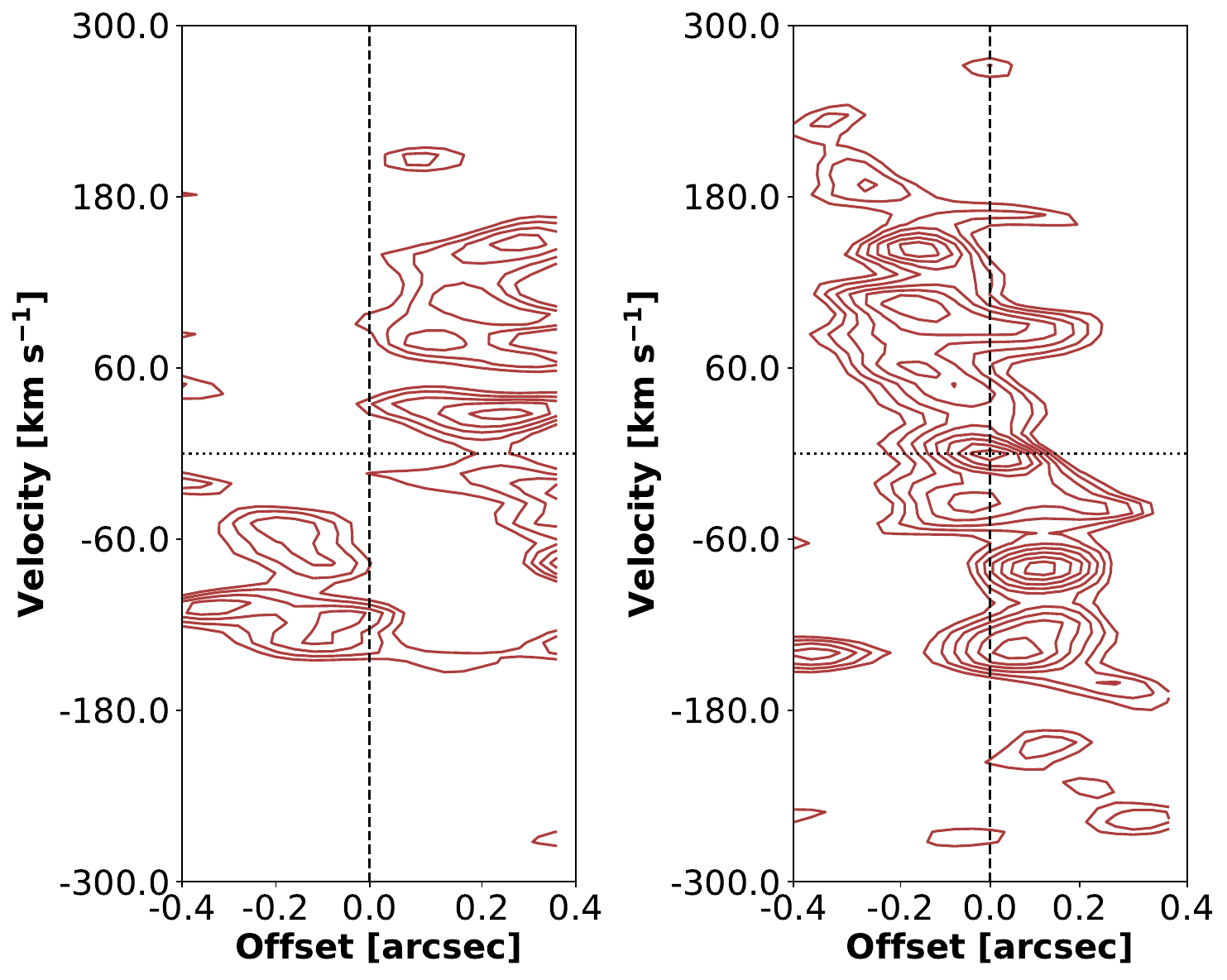}

\caption{\hi 21 cm PV plots. \textit{Left}: Cut B. \textit{Right}: Cut A. The cuts are shown as dashed blue (B) and red (A) lines in the bottom-left panel of Fig.~\ref{fig:21cm_mom_maps}. The vertical dashed line represents the midpoint of the respective cuts, and the horizontal dotted line marks zero velocity relative to $z_{\rm core}$. The contours are at levels of $-1.1$ mJy beam$^{-1}$ $\times$ (2.0, 2.5, 3.0, 3.5, 4.0, 4.5, 5.0, 5.5, 6.0, 6.5, 7.0, 7.5, 8.0), where 1.1 mJy beam$^{-1}$ is the RMS noise in the image.}
\label{fig:h1_pv_plot}
\end{figure}

\section{Discussion}

\subsection{Discs in merger remnants}

Earlier numerical simulations predicted that major mergers of two disc galaxies result in the formation of a spheroid-dominated early-type galaxy \citep[][]{barnes1992, naab2003}. Contrary to this scenario, more recent high-resolution simulations have shown that not all major mergers become early-type galaxies, but some emerge as disc-dominated late-type galaxies \citep[][]{hopkins2009, xu2010}. During the merger, a portion of the gas that does not lose a significant amount of angular momentum survives the collision, reforming into a gaseous disc \citep[][]{springel_her2005}. The gas that falls to the centre of the galaxy could contribute to the formation of a spheroidal component and to a starburst in the nuclear regions.

From CO observations, \citet[][]{ueda2014} found that 24 out of 30 merger remnants in their sample have kinematical signatures of rotating molecular gas discs, including nuclear rings. The sizes of these discs are in the range of 1--9 kpc. The current system, TXS 1033+026, is an example of a merger remnant with a kiloparsec-scale $\rm H\alpha$ disc. All these results agree with the simulations of \citet[][]{springel_her2005, hopkins2009}.

The double-peaked $\rm H\alpha$ lines detected in SALT spectroscopy towards TXS 1033+026 \citep[see][]{dutta2018} were thought to arise from the nuclear surroundings of two merging galaxies. Recently, however, using data from the Mapping Nearby Galaxies at Apache Point Observatory (MaNGA) survey \citep[][]{bundy2015}, \citet[][]{zhang2025} found that double $\rm H\alpha$ emission-line peaks tend to arise from circumnuclear discs rather than from double-AGN systems, as previously thought. The current system is consistent with this study as well; the Gemini IFU observations reveal that the lines likely arise from red- and blueshifted components of a circumnuclear kiloparsec-scale disc.

\subsection{Counter-rotating discs}

One of the forms of counter-rotation in galaxies is gas versus gas, where two gaseous discs counter-rotate with respect to each other \citep[e.g.][]{corsini2014}. Counter-rotating systems are detected by looking for signatures of opposite orientation of gas emission-line velocities in two-dimensional spectra \citep[e.g.][]{galletta1987} or in PV diagrams \citep[e.g.][]{bureau2006, corsini2014}. Counter-rotating gaseous components are direct evidence of past gas accretion, because a system that is evolving through self-gravitation cannot form a component with counter-rotating angular momentum \citep[][]{beom2022}. The formation process of counter-rotating components results in a loss of angular momentum due to hydrodynamic friction \citep[][]{beom2022}. Due to counter-rotation, the colliding gas streams cancel out angular momentum, forcing the material inwards. Counter-rotating systems are therefore good cases for studying the process of AGN fuelling and gas accretion.

Arp 220 \citep[][]{sakamoto1999}, NGC 0253 \citep[][]{zhao2001}, NGC 3593 \citep[][]{burillo2000}, Centaurus A (NGC 5128; \citealt[][]{espada2009, morganti2010}), and NGC 7332 \citep[][]{fisher1994, plana1996} are some of the known cases of counter-rotating circumnuclear gas discs. The discs are found to be rotating in an orientation opposite to that of a larger kiloparsec-scale disc or the galactic disc. In NGC 5128, \citet[][]{espada2009} additionally find that the central molecular disc is likely warped. NGC 7332 is found to have two discs of ionised gas rotating in opposite senses to each other. The gas displays non-circular motions, indicating that it has not reached equilibrium \citep[][]{fisher1994, plana1996}. Arp 220 is a complex merging system with two nuclei, each of which has a circumnuclear gas disc of size $\approx 100$ pc, and these discs are counter-rotating with respect to each other. Both nuclei are embedded in a larger $\rm \approx 1 \ kpc$ rotating molecular gas disc, creating a complex gas system.

The current system, TXS 1033+026, is a case of counter-rotation of ionised ($\rm H\alpha$) and neutral gas (\hi). However, the \hi plots show a more complex velocity structure than a simple circular rotation, along with indications of a warp, suggesting that the gas in the central regions has not yet settled into equilibrium.

\subsection{Covering fraction}

In Fig.~\ref{fig:tot_h1_spec} we compare the \hi 21 cm spectra from the GMRT and eMERLIN observations. The eMERLIN spectrum is extracted against the complete region showing absorption in Fig.~\ref{fig:21cm_mom_maps}. Note that, in the maps, the gas does not cover the entire background continuum emission. The eMERLIN and GMRT spectra agree with each other within the noise, indicating that most of the GMRT absorption is likely recovered in the eMERLIN data. However, we cannot rule out that a small fraction of absorption, not recovered in the spectrum, could arise against the remaining background continuum emission.

The covering fraction is defined as the ratio of the flux density of the emission covered by the absorbing gas to the flux density of the total emission \citep[][]{briggs1983}. The integrated flux density of the emission covered by the gas in Fig.~\ref{fig:21cm_mom_maps} is $90$\,mJy. The GMRT spectrum of \citet[][]{dutta2018} was extracted at the peak of the core, which has a flux density of $132$\,mJy. This yields a covering fraction of 0.7 \citep[see e.g.][]{curran2013b}. \citet[][]{dutta2018} reported a VOD of $\rm 50 \pm 1 \ km \ s^{-1}$ from the GMRT spectrum. Using the eMERLIN spectrum (Fig.~\ref{fig:tot_h1_spec}), with a background flux density of $90$\,mJy, we estimate a VOD of $\rm 90 \pm 3\ km \ s^{-1}$ between the nulls of the absorption. This is the continuum-weighted VOD over the absorbing region.

\section{Conclusions}

We conducted eMERLIN and Gemini IFU observations towards the target TXS 1033+026 to trace \hi 21 cm absorption and $\rm H\alpha$ emission at sub-kiloparsec scales. These are the first results showing resolved $\rm H\alpha$ and \hi 21 cm structures towards this target. From our observations, we deduce the following:
\begin{itemize}
    \item We find a symmetric, two-sided core-jet structure for the radio source based on L- and C-band radio continuum images. Multiple peaks in the L band, other than the core, indicate a double-double radio AGN morphology. The direction of jet propagation aligns with kiloparsec-scale radio lobes detected earlier in FIRST images, indicating a possibility of multiple episodes of jet activity.

    \item We estimate a redshift of $z_{\rm core} = 0.05012 \pm 0.00011$ by extracting the optical spectrum at the location of the radio peak corresponding to the AGN core.
   \item The $\rm H\alpha$ maps suggest the presence of a kiloparsec-scale circumnuclear gas disc, rotating around the radio core. The disc shows signs of a warp.
   \item
   The results from \hi 21 cm observations reveal a complex gaseous structure, with a gas component that is likely counter-rotating with respect to the $\rm H\alpha$ disc. The component shows indications of a warp. Another component of the structure appears to represent a gas ring that is counter-rotating relative to the first one, and co-rotating relative to the $\rm H\alpha$ disc. However, the total structure could have more complex kinematics than our interpretation suggests.
   \item
   Using the flux density of the emission covered by the absorbing gas, we estimate a continuum-weighted VOD of $\rm 90 \pm 3 \ km \ s^{-1}$ for the \hi 21 cm absorption over the absorbing region.

\end{itemize}
In a subsequent article, we will discuss the kinematics and other physical properties of the circumnuclear gas using $\rm H\alpha$, $\rm H\beta$, Na\,{\sc i}, [O\,{\sc i}], [N\,{\sc ii}], and [S\,{\sc ii}] lines detected in the Gemini observations.

\section*{Data availability}
{\raggedright
The Gemini IFU data are available at the Gemini Observatory Archive, which can be accessed with the link \url{https://archive.gemini.edu/searchform/GN-2025A-FT-105}. The eMERLIN data are available at \url{https://www.e-merlin.ac.uk/distribute/CY14/CY14211/CY14211.html}. The reduced images, PV diagrams, and spectra shown in Figs.~\ref{fig:21cm_cont_maps}--\ref{fig:h1_pv_plot}, \ref{fig:core_optical_spec}, \ref{fig:ha_mom2}, and \ref{fig:h1_spec} are only available in electronic form at the CDS via anonymous ftp to \url{cdsarc.u-strasbg.fr} (130.79.128.5) or via \url{http://cdsweb.u-strasbg.fr/cgi-bin/qcat?J/A+A/}.\par}

\begin{acknowledgements}
This work is partly supported by the National SKA Program of China (No. 2025SKA0130100).
JA thanks Nissim Kanekar for discussions and several inputs. RZS acknowledges the support from the National Natural Science Foundation of China (Grant No.\ 12603034), the China Postdoctoral Science Foundation (Grant No.\ 2025M783232), and the Shanghai Super Postdoctoral Incentive Program (No.\ 2025322). HY is supported by the National Research Foundation of Korea (NRF) grant funded by the Korea government (MSIT) (RS-2025-00516062).
XLY is supported by the National SKA Program of China (Grant No. 2025SKA0130100) and the National Natural Science Foundation of China (Key Program) (Grant No. 12233005).
Analysis in this paper is based on observations obtained under project GN-2025A-FT-105 (P.I. Yoon) at the international Gemini Observatory, a program of NSF NOIRLab, which is managed by the Association of Universities for Research in Astronomy (AURA) under a cooperative agreement with the U.S. National Science Foundation on behalf of the Gemini Observatory partnership: the U.S. National Science Foundation
(United States), National Research Council (Canada), Agencia Nacional de Investigación y Desarrollo (Chile), Ministerio de Ciencia, Tecnología e Innovación (Argentina), Ministério da Ciência, Tecnologia, Inovações e Comunicações (Brazil), and Korea Astronomy and Space Science Institute (Republic of
Korea).

\end{acknowledgements}

\bibliographystyle{aa}

\bibliography{aa_example}

\begin{thebibliography}{41}
\expandafter\ifx\csname natexlab\endcsname\relax\def\natexlab#1{#1}\fi

\bibitem[{{Abolfathi} {et~al.}(2017){Abolfathi}, {Aguado}, {Aguilar}, {Allende
  Prieto}, {Almeida}, {Tasnim Ananna}, {Anders}, {Anderson}, {Andrews},
  {Anguiano}, \& et~al.}]{abolfathi2017}
{Abolfathi}, B., {Aguado}, D.~S., {Aguilar}, G., {et~al.} 2017, ArXiv e-prints
  [\eprint[arXiv]{1707.09322}]

\bibitem[{{Barnes} \& {Hernquist}(1992)}]{barnes1992}
{Barnes}, J.~E. \& {Hernquist}, L. 1992, \araa, 30, 705

\bibitem[{{Bekki}(1995)}]{bekki1995}
{Bekki}, K. 1995, \mnras, 276, 9

\bibitem[{{Beom} {et~al.}(2022){Beom}, {Bizyaev}, {Walterbos}, \&
  {Chen}}]{beom2022}
{Beom}, M., {Bizyaev}, D., {Walterbos}, R. A.~M., \& {Chen}, Y. 2022, \mnras,
  516, 3175

\bibitem[{{Best} \& {Heckman}(2012)}]{best2012}
{Best}, P.~N. \& {Heckman}, T.~M. 2012, \mnras, 421, 1569

\bibitem[{{Briggs} \& {Wolfe}(1983)}]{briggs1983}
{Briggs}, F.~H. \& {Wolfe}, A.~M. 1983, \apj, 268, 76

\bibitem[{{Bundy} {et~al.}(2015){Bundy}, {Bershady}, {Law}, {Yan}, {Drory},
  {MacDonald}, {Wake}, {Cherinka}, {S{\'a}nchez-Gallego}, {Weijmans}, {Thomas},
  {Tremonti}, {Masters}, {Coccato}, {Diamond-Stanic}, {Arag{\'o}n-Salamanca},
  {Avila-Reese}, {Badenes}, {Falc{\'o}n-Barroso}, {Belfiore}, {Bizyaev},
  {Blanc}, {Bland-Hawthorn}, {Blanton}, {Brownstein}, {Byler}, {Cappellari},
  {Conroy}, {Dutton}, {Emsellem}, {Etherington}, {Frinchaboy}, {Fu}, {Gunn},
  {Harding}, {Johnston}, {Kauffmann}, {Kinemuchi}, {Klaene}, {Knapen},
  {Leauthaud}, {Li}, {Lin}, {Maiolino}, {Malanushenko}, {Malanushenko}, {Mao},
  {Maraston}, {McDermid}, {Merrifield}, {Nichol}, {Oravetz}, {Pan}, {Parejko},
  {Sanchez}, {Schlegel}, {Simmons}, {Steele}, {Steinmetz}, {Thanjavur},
  {Thompson}, {Tinker}, {van den Bosch}, {Westfall}, {Wilkinson}, {Wright},
  {Xiao}, \& {Zhang}}]{bundy2015}
{Bundy}, K., {Bershady}, M.~A., {Law}, D.~R., {et~al.} 2015, \apj, 798, 7

\bibitem[{{Bureau} \& {Chung}(2006)}]{bureau2006}
{Bureau}, M. \& {Chung}, A. 2006, \mnras, 366, 182

\bibitem[{{Calabr{\`o}} {et~al.}(2019){Calabr{\`o}}, {Daddi}, {Fensch},
  {Bournaud}, {Cibinel}, {Puglisi}, {Jin}, {Delvecchio}, \&
  {D'Eugenio}}]{calabro2019}
{Calabr{\`o}}, A., {Daddi}, E., {Fensch}, J., {et~al.} 2019, \aap, 632, A98

\bibitem[{{Corsini}(2014)}]{corsini2014}
{Corsini}, E.~M. 2014, in Astronomical Society of the Pacific Conference
  Series, Vol. 486, Multi-Spin Galaxies, ed. E.~{Iodice} \& E.~M. {Corsini}, 51

\bibitem[{{Curran} {et~al.}(2013){Curran}, {Allison}, {Glowacki}, {Whiting}, \&
  {Sadler}}]{curran2013b}
{Curran}, S.~J., {Allison}, J.~R., {Glowacki}, M., {Whiting}, M.~T., \&
  {Sadler}, E.~M. 2013, \mnras, 431, 3408

\bibitem[{{Durr{\'e}} \& {Mould}(2019)}]{durre2019}
{Durr{\'e}}, M. \& {Mould}, J. 2019, \apj, 870, 37

\bibitem[{{Dutta} {et~al.}(2018){Dutta}, {Srianand}, \& {Gupta}}]{dutta2018}
{Dutta}, R., {Srianand}, R., \& {Gupta}, N. 2018, \mnras, 480, 947

\bibitem[{{Ellison} {et~al.}(2013){Ellison}, {Mendel}, {Patton}, \&
  {Scudder}}]{ellison2013}
{Ellison}, S.~L., {Mendel}, J.~T., {Patton}, D.~R., \& {Scudder}, J.~M. 2013,
  \mnras, 435, 3627

\bibitem[{{Espada} {et~al.}(2009){Espada}, {Matsushita}, {Peck}, {Henkel},
  {Iono}, {Israel}, {Muller}, {Petitpas}, {Pihlstr{\"o}m}, {Taylor}, \&
  {Dinh-V-Trung}}]{espada2009}
{Espada}, D., {Matsushita}, S., {Peck}, A., {et~al.} 2009, \apj, 695, 116

\bibitem[{{Fisher} {et~al.}(1994){Fisher}, {Illingworth}, \&
  {Franx}}]{fisher1994}
{Fisher}, D., {Illingworth}, G., \& {Franx}, M. 1994, \aj, 107, 160

\bibitem[{{Freitas} {et~al.}(2018){Freitas}, {Riffel}, {Storchi-Bergmann},
  {Elvis}, {Robinson}, {Crenshaw}, {Nagar}, {Lena}, {Schmitt}, \&
  {Kraemer}}]{freitas2018}
{Freitas}, I.~C., {Riffel}, R.~A., {Storchi-Bergmann}, T., {et~al.} 2018,
  \mnras, 476, 2760

\bibitem[{{Galletta}(1987)}]{galletta1987}
{Galletta}, G. 1987, \apj, 318, 531

\bibitem[{{Garc{\'\i}a-Burillo} {et~al.}(2000){Garc{\'\i}a-Burillo}, {Sempere},
  {Combes}, {Hunt}, \& {Neri}}]{burillo2000}
{Garc{\'\i}a-Burillo}, S., {Sempere}, M.~J., {Combes}, F., {Hunt}, L.~K., \&
  {Neri}, R. 2000, \aap, 363, 869

\bibitem[{{Gaspari} {et~al.}(2013){Gaspari}, {Ruszkowski}, \&
  {Oh}}]{gaspari2013}
{Gaspari}, M., {Ruszkowski}, M., \& {Oh}, S.~P. 2013, \mnras, 432, 3401

\bibitem[{{Hopkins} {et~al.}(2013){Hopkins}, {Cox}, {Hernquist}, {Narayanan},
  {Hayward}, \& {Murray}}]{hopkins2013}
{Hopkins}, P.~F., {Cox}, T.~J., {Hernquist}, L., {et~al.} 2013, \mnras, 430,
  1901

\bibitem[{{Hopkins} {et~al.}(2009){Hopkins}, {Cox}, {Younger}, \&
  {Hernquist}}]{hopkins2009}
{Hopkins}, P.~F., {Cox}, T.~J., {Younger}, J.~D., \& {Hernquist}, L. 2009,
  \apj, 691, 1168

\bibitem[{{Kilerci Eser} {et~al.}(2014){Kilerci Eser}, {Goto}, \&
  {Doi}}]{kilerci2014}
{Kilerci Eser}, E., {Goto}, T., \& {Doi}, Y. 2014, \apj, 797, 54

\bibitem[{{Knapen} {et~al.}(2004){Knapen}, {Whyte}, {de Blok}, \& {van der
  Hulst}}]{knapen2004}
{Knapen}, J.~H., {Whyte}, L.~F., {de Blok}, W.~J.~G., \& {van der Hulst}, J.~M.
  2004, \aap, 423, 481

\bibitem[{{Kolcu} {et~al.}(2023){Kolcu}, {Maciejewski}, {Gadotti}, {Fragkoudi},
  {Erwin}, {S{\'a}nchez-Bl{\'a}zquez}, {Neumann}, {Van de Ven}, {de
  S{\'a}-Freitas}, {Longmore}, \& {Debattista}}]{kolcu2023}
{Kolcu}, T., {Maciejewski}, W., {Gadotti}, D.~A., {et~al.} 2023, \mnras, 524,
  207

\bibitem[{{Marecki} {et~al.}(2006){Marecki}, {Thomasson}, {Mack}, \&
  {Kunert-Bajraszewska}}]{marecki2006}
{Marecki}, A., {Thomasson}, P., {Mack}, K.-H., \& {Kunert-Bajraszewska}, M.
  2006, \aap, 448, 479

\bibitem[{{Morganti}(2010)}]{morganti2010}
{Morganti}, R. 2010, \pasa, 27, 463

\bibitem[{{Morganti} {et~al.}(2023){Morganti}, {Murthy}, {Oosterloo},
  {Blanchard}, {Cook}, {Paragi}, {Orienti}, {Nagai}, \&
  {Schulz}}]{morganti2023}
{Morganti}, R., {Murthy}, S., {Oosterloo}, T., {et~al.} 2023, \aap, 678, A42

\bibitem[{{Murthy} {et~al.}(2024){Murthy}, {Morganti}, {Oosterloo}, {Schulz},
  \& {Paragi}}]{murthy2024}
{Murthy}, S., {Morganti}, R., {Oosterloo}, T., {Schulz}, R., \& {Paragi}, Z.
  2024, \aap, 688, A84

\bibitem[{{Naab} \& {Burkert}(2003)}]{naab2003}
{Naab}, T. \& {Burkert}, A. 2003, \apj, 597, 893

\bibitem[{{Plana} \& {Boulesteix}(1996)}]{plana1996}
{Plana}, H. \& {Boulesteix}, J. 1996, \aap, 307, 391

\bibitem[{{Rao} {et~al.}(2023){Rao}, {Kharb}, {Rubinur}, {Silpa}, {Roy},
  {Sebastian}, {Singh}, {Baghel}, {Manna}, \& {Ishwara-Chandra}}]{rao2023}
{Rao}, V.~V., {Kharb}, P., {Rubinur}, K., {et~al.} 2023, \mnras, 524, 1615

\bibitem[{{Saikia} \& {Jamrozy}(2009)}]{saikia2009}
{Saikia}, D.~J. \& {Jamrozy}, M. 2009, Bulletin of the Astronomical Society of
  India, 37, 63

\bibitem[{{Sakamoto} {et~al.}(1999){Sakamoto}, {Scoville}, {Yun}, {Crosas},
  {Genzel}, \& {Tacconi}}]{sakamoto1999}
{Sakamoto}, K., {Scoville}, N.~Z., {Yun}, M.~S., {et~al.} 1999, \apj, 514, 68

\bibitem[{{Springel} \& {Hernquist}(2005)}]{springel_her2005}
{Springel}, V. \& {Hernquist}, L. 2005, \apjl, 622, L9

\bibitem[{{Ueda} {et~al.}(2014){Ueda}, {Iono}, {Yun}, {Crocker}, {Narayanan},
  {Komugi}, {Espada}, {Hatsukade}, {Kaneko}, {Matsuda}, {Tamura}, {Wilner},
  {Kawabe}, \& {Pan}}]{ueda2014}
{Ueda}, J., {Iono}, D., {Yun}, M.~S., {et~al.} 2014, \apjs, 214, 1

\bibitem[{{White} {et~al.}(1997){White}, {Becker}, {Helfand}, \&
  {Gregg}}]{white1997}
{White}, R.~L., {Becker}, R.~H., {Helfand}, D.~J., \& {Gregg}, M.~D. 1997,
  \apj, 475, 479

\bibitem[{{Xu} {et~al.}(2010){Xu}, {Narayanan}, \& {Walker}}]{xu2010}
{Xu}, X., {Narayanan}, D., \& {Walker}, C. 2010, \apjl, 721, L112

\bibitem[{{Zhang} {et~al.}(2025){Zhang}, {Chen}, {Shen}, {Kauffmann}, {Bao},
  {Zhou}, {Jin}, \& {Zhou}}]{zhang2025}
{Zhang}, Z., {Chen}, Y., {Shen}, S., {et~al.} 2025, \mnras, 543, 3089

\bibitem[{{Zhao} {et~al.}(2001){Zhao}, {Goss}, {Ulvestad}, \&
  {Anantharamaiah}}]{zhao2001}
{Zhao}, J., {Goss}, W.~M., {Ulvestad}, J.~S., \& {Anantharamaiah}, K.~R. 2001,
  in Astronomical Society of the Pacific Conference Series, Vol. 240, Gas and
  Galaxy Evolution, ed. J.~E. {Hibbard}, M.~{Rupen}, \& J.~H. {van Gorkom}, 404

\bibitem[{{Zhou} {et~al.}(2017){Zhou}, {Federrath}, {Yuan}, {Bian}, {Medling},
  {Shi}, {Bland-Hawthorn}, {Bryant}, {Brough}, {Catinella}, {Croom}, {Goodwin},
  {Goldstein}, {Green}, {Konstantopoulos}, {Lawrence}, {Owers}, {Richards}, \&
  {Sanchez}}]{zhou2017}
{Zhou}, L., {Federrath}, C., {Yuan}, T., {et~al.} 2017, \mnras, 470, 4573

\end{thebibliography}

\begin{appendix}

\section{Additional figures} \label{app:figures}

\begin{figure}[!ht]

\includegraphics[width=0.5\textwidth]{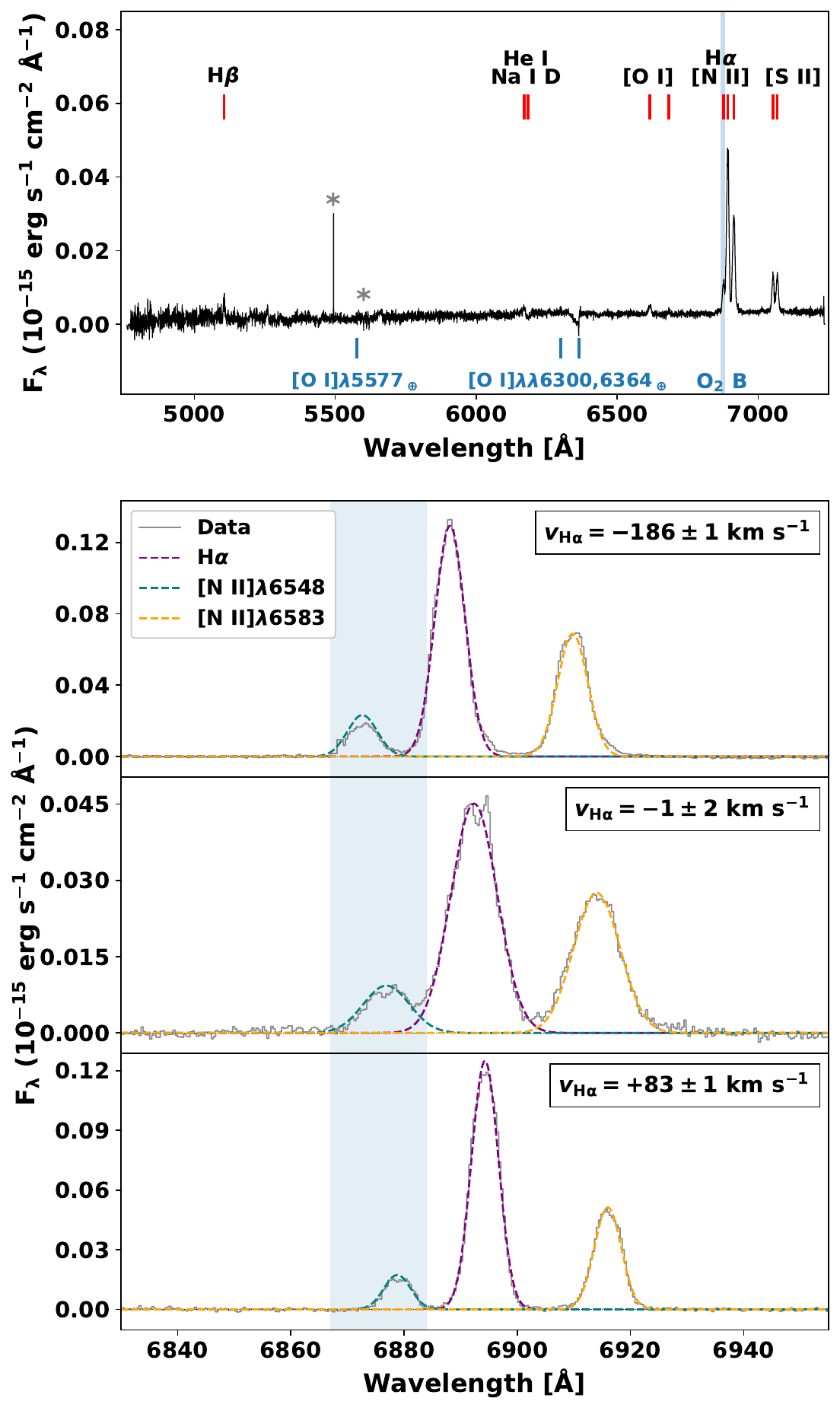}
\caption{\textit{Top}: Gemini optical spectrum extracted at the location of the radio core in the eMERLIN C-band image. The red markers indicate the detected emission and absorption lines: H$\beta$, He\,{\sc i}\,$\lambda$5876 and Na\,{\sc i}\,D, [O\,{\sc i}]\,$\lambda\lambda$6300,6364, H$\alpha$ and [N\,{\sc ii}]\,$\lambda\lambda$6548,6583, and [S\,{\sc ii}]\,$\lambda\lambda$6716,6731. The blue markers indicate the night-sky [O\,{\sc i}]\,$\lambda$5577 and $\lambda\lambda$6300,6364 lines, and the shaded region marks the telluric O$_2$ B band (6867--6884\,\AA). The asterisks mark cosmic-ray residuals. \textit{Bottom}: Tied triple-Gaussian fits to the H$\alpha$ and [N\,{\sc ii}]\,$\lambda\lambda$6548,6583 lines at the three locations shown in Fig.~\ref{fig:ha_mom2}, from top to bottom. A fitted linear continuum has been subtracted from each spectrum. The box in each panel gives the heliocentric H$\alpha$ line-of-sight velocity relative to $z_{\rm core}$. The shaded region marks the telluric O$_2$ B band, which overlaps the redshifted [N\,{\sc ii}]\,$\lambda$6548 line. As the amplitude of this line is tied to that of [N\,{\sc ii}]\,$\lambda$6583, excluding the B band from the fit changes the fitted velocities and velocity dispersions by $\leq 2$\,km\,s$^{-1}$.
 } \label{fig:core_optical_spec}
\end{figure}

\begin{figure}[ht]

\includegraphics[width=0.5\textwidth]{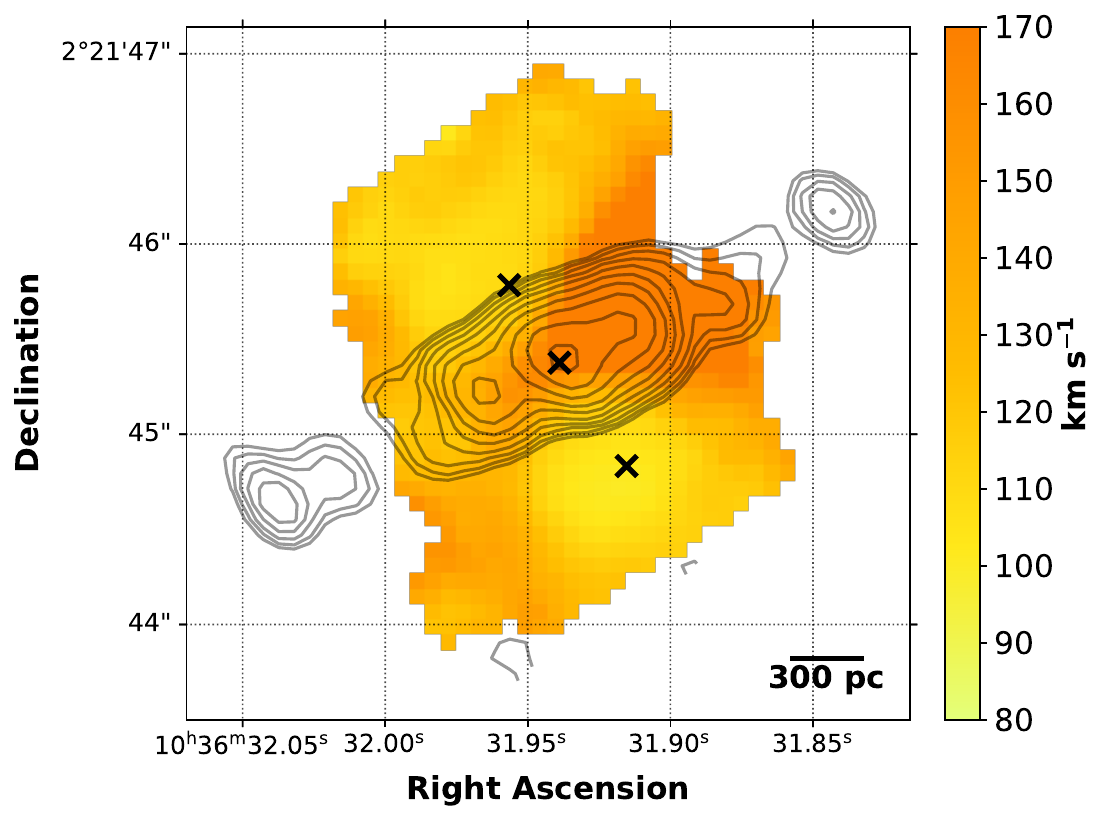}

\caption{$\rm H\alpha$ velocity dispersion plot. The black contours represent the eMERLIN L-band continuum emission, at the same levels as those in Fig.~\ref{fig:21cm_cont_maps}. The black markers represent the positions where the spectra in the lower panels of Fig.~\ref{fig:core_optical_spec} are extracted.}\label{fig:ha_mom2}
\end{figure}

\begin{figure*}[ht]

\includegraphics[width=0.97\textwidth]{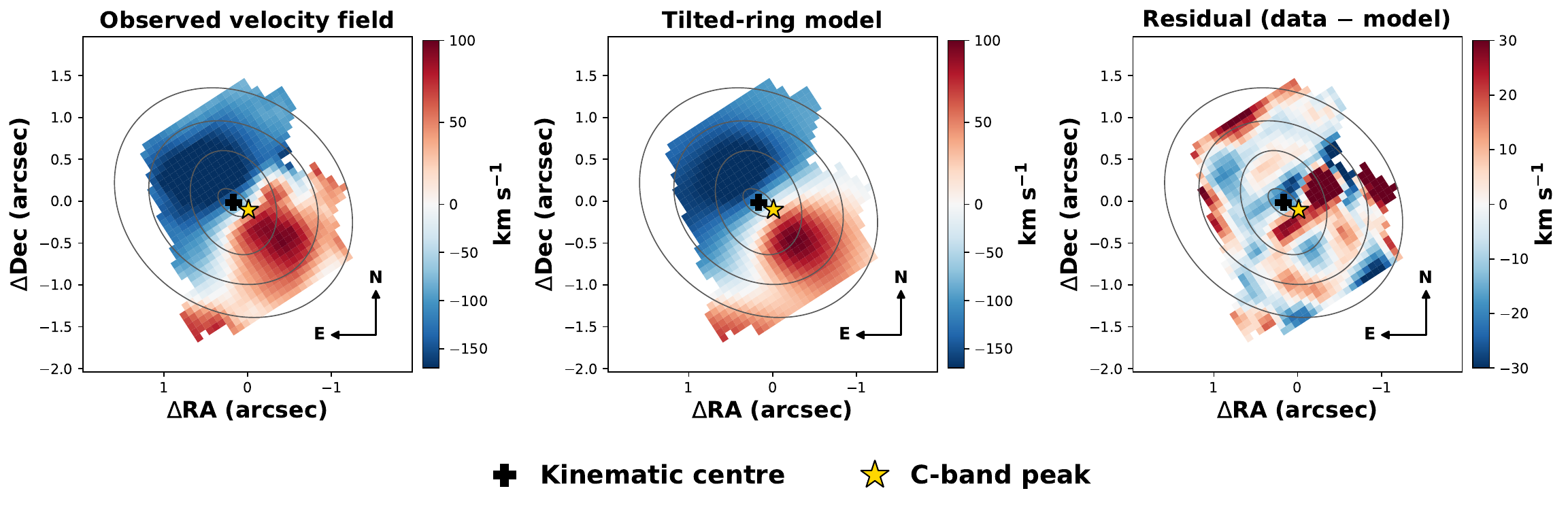}

\caption{\textit{Left}: Observed velocity field of the $\rm H\alpha$ line. \textit{Middle}: Model velocity field. \textit{Right}: Residual velocities (see Sect.~\ref{sec:model_fit}). The black lines represent the sky-projected ellipses corresponding to each ring's mid-radius, drawn with the fitted position angle and inclination of each ring (Table~\ref{tab:ringfit}). The kinematic centre of the rings lies within 0.2$''$ of the C-band radio peak.} \label{fig:ring_fits}
\end{figure*}

\begin{table}
\caption{Parameters of the tilted rings fitted to the H$\alpha$ velocity field.}
\label{tab:ringfit}
\centering
\begin{tabular}{cccc}
\hline\hline
Ring & Mid-radius & PA & Inc. \\
     & ($''$)     & ($^\circ$) & ($^\circ$) \\
\hline
1 & 0.22 & 231 & 55 \\
2 & 0.65 & 205 & 42 \\
3 & 1.08 & 230 & 34 \\
4 & 1.51 & 231 & 33 \\
\hline
\end{tabular}
\tablefoot{The columns are: the ring number, the mid-radius of the ring, the position angle of the receding (redshifted) major axis, measured from north to east, and the inclination. All rings share a common fitted kinematic centre at RA = 10:36:31.948, Dec = +02:21:45.47 (J2000; $\pm0.6''$).}
\end{table}

\begin{figure*}[ht]

\centering
\includegraphics[width=0.32\textwidth]{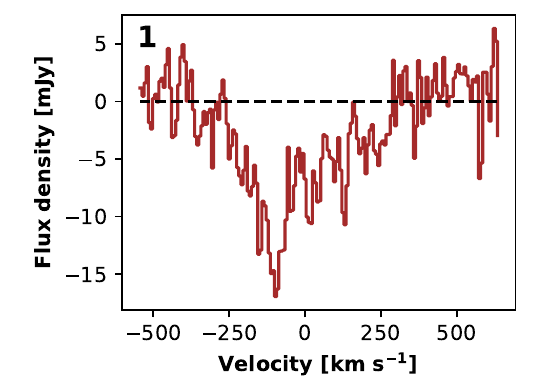}
\includegraphics[width=0.32\textwidth]{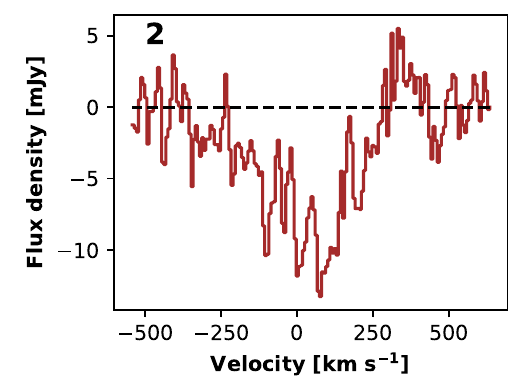}
\includegraphics[width=0.32\textwidth]{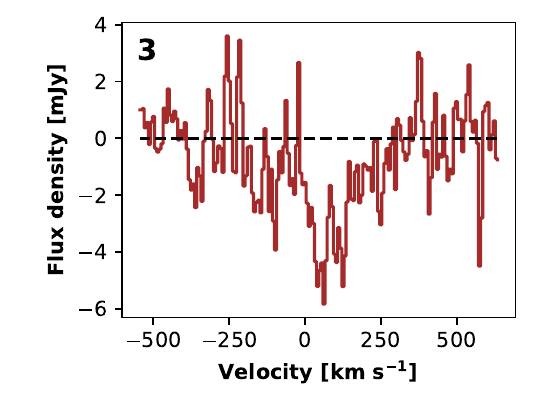}\\[2mm]
\includegraphics[width=0.32\textwidth]{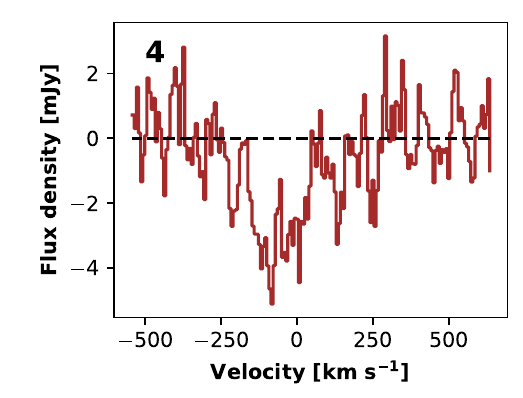}
\includegraphics[width=0.32\textwidth]{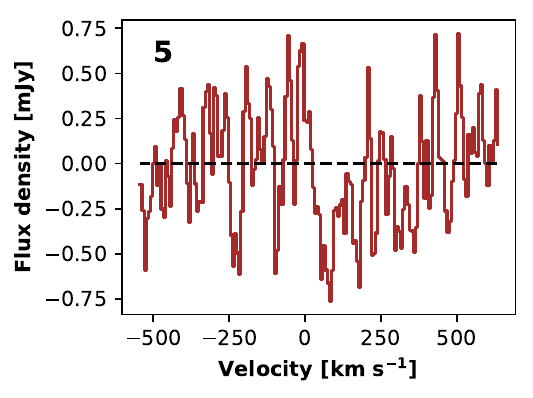}
\includegraphics[width=0.32\textwidth]{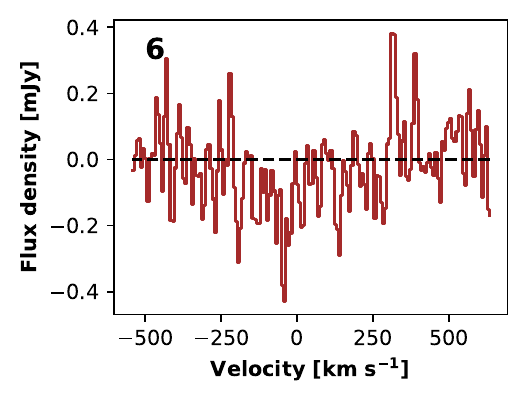}

\caption{\hi 21 cm absorption spectra from eMERLIN observations. The labels represent the regions marked in the bottom-left panel of Fig.~\ref{fig:21cm_mom_maps}.} \label{fig:h1_spec}
\end{figure*}

\FloatBarrier

\end{appendix}

\end{document}